\documentclass[conference]{IEEEtran}
\usepackage[utf8]{inputenc}
\usepackage{graphicx} 
\usepackage{float}
\ifCLASSINFOpdf
\else
\fi
\usepackage[cmex10]{amsmath}
\usepackage{url}
\begin{document}

\title{Evaluation of a RGB-LED-based Emotion Display for Affective Agents}

\author{\IEEEauthorblockN{Johannes Feldmaier, Tamara Marmat, Johannes Kuhn, and Klaus Diepold}
\IEEEauthorblockA{Department of Electrical, Electronic and Computer Engineering\\
Technische Universität München\\
Munich, Germany\\}
\IEEEauthorblockA{Email: johannes.feldmaier@tum.de, tamara.marmat@tum.com, johannes.kuhn@tum.de, klaus.diepold@tum.de\\}}

% use for special paper notices
%\IEEEspecialpapernotice{(Invited Paper)}

% make the title area
\maketitle

\begin{abstract}
%\boldmath

% IEEEtran.cls defaults to using nonbold math in the Abstract.
% This preserves the distinction between vectors and scalars. However,
% if the conference you are submitting to favors bold math in the abstract,
% then you can use LaTeX's standard command \boldmath at the very start
% of the abstract to achieve this. Many IEEE journals/conferences frown on
% math in the abstract anyway.

% no keywords
Technology has become an essential part in every aspect of our lives. However the key 
to a successful implementation of a technology depends on the acceptance by the 
general public. In order to increase the acceptance various approaches can be applied. 
In this paper, we will examine the human-robot emotional interaction by investigating the 
capabilities of a developed low-resolution RGB-LED display in the context of artificial 
emotions. We are focusing on four of the most representative human emotions which include 
happiness, anger, sadness and fear. We will work with colors and dynamic light patterns 
which are supposed to evoke various associations. In an experiment, the use these patterns 
as expressions of emotions are validated. The results of the conducted study show that 
some of the considered basic emotions can be recognized by human observers.  

\end{abstract}
% For peer review papers, you can put extra information on the cover
% page as needed:
% \ifCLASSOPTIONpeerreview
% \begin{center} \bfseries EDICS Category: 3-BBND \end{center}
% \fi
%
% For peerreview papers, this IEEEtran command inserts a page break and
% creates the second title. It will be ignored for other modes.
\IEEEpeerreviewmaketitle

\section{Introduction}
% no \IEEEPARstart
Emotions are a highly complex area and still subject of research. Our behavior is influenced 
by emotions and our previous experiences. Emotions appear to serve several physical and 
psychological purposes. The impact of emotions can be observed in many different situations, 
like the emotion fear which leads to an increased heart rate. In case of a frightening 
situation, emotions have the function of avoiding an event which could be potentially 
dangerous for us. However, a lot of emotions emerge spontaneously and are unconsciously 
conveyed via nonverbal behavior \cite{terada2012artificial}. Regardless of the question if it will 
ever be possible to model emotions completely and consequently develop a machine with true 
emotions, expressing emotions in artificial intelligence could be useful. For example it 
could increase friendliness and thus increase acceptance of new technologies. However it 
also could be used to influence people, either in a positive or a negative way. It is an 
undisputed fact that robots will become part of our daily life. As robots grow smarter, they 
can undertake much more complex tasks. Thus, the human-robot interaction, as part of our 
daily life, should be designed as comfortable as possible. In human-to-human communication, 
emotions are conveyed via body language especially facial expression. Communicating 
with emotions means a fast and natural way of non-verbal communication. Studies show 
that two-thirds of communication between humans is non-verbal \cite{mehrabian1977nonverbal, 
hogan2003can, mavridis2015commun}. Non-verbal communication involves cues (mostly visual, 
like eye contact or facial expressions) that depend on the affective state of the communication 
partner and are inherently perceived and interpreted by a human observer. The cues are to 
some some extent language- and culture-independent. Robots equipped with facial features 
\cite{fong2003survey}, or complete androids, like the Geminoid F \cite{asano2011facial}, 
are used to study affective expressions. But also appearance-constrained robots, which do not 
have facial features can enhance communication with non-verbal affective expressions, cf. 
\cite{bethel2008survey, bethel2009prelim, mavridis2015commun}.

The non-verbal way of communication using artificial emotions requires that those emotions 
are modeled correctly within robotic systems. Computational modeling of emotions is an active 
area of research. An overview of those models can be found in \cite{calvo2015handbook}. Going one step 
further, whole body emotional expressions like the one of the robot KOBIAN \cite{zecca2008design} 
were already developed and the expression recognition rates checked. In contrast to KOBIAN, 
simpler methods to achieve a affective expression are required, in order to cut down the 
costs and the implementation expense.

Studies on using colored light on a robot's face were conducted. Depending on the emotion, 
either the head or the cheeks where illuminated with the appropriate color. For example, 
a red illuminated head represents an angry emotional state. Studies on facial colors on a 
humanoid robot have confirmed that e.g. yellow represents joy and blue represents sadness 
\cite{matsui2010model}.  

A method of expressing emotions by dynamically changing the color and luminosity of the 
robot's body, was already proposed. The parameters, like the hue value, the frequencies, 
and the type of the blinking waveforms were selected by the participants. For example, intense 
emotions were represented by a rectangular waveform with a high frequency. The impact of 
specific color combinations and color patterns was not considered \cite{terada2012artificial}.

Also, a study on the expression of emotions using lights sources integrated into apparel 
was done. Here, the type of the used lights and dynamic pattern (blink speed and blink rhythm) 
are used as characteristic visual languages. However, a user study to prove the proposed 
matrix for expressing different emotions using light in cloths was not conducted 
\cite{choi2007study}. 

In order to map emotions to specific colors we first need to understand the emotion model. 
To determine the relationship between colors and emotions, we rely on several color-emotion 
theories/models \cite{nijdam2009mapping}. In the following, the application of these theories 
in order to evoke an emotional state in an human observer are investigated. A study is 
performed to determine how good the emotions can be recognized by an observer. 

As already mentioned, we focus on a sub-set of the emotional states: happiness, anger, fear 
and sadness. These four distinct emotions are presented by four specific and dynamic light patterns 
on our developed display. After each pattern is displayed the observer answers a short questionnaire, 
to evaluate whether the emotional state of the robot was recognized. Since the description of an 
emotion is difficult to analyze, at the beginning of the study a calibration test is performed. 
This test consists of situations where the general answer of a human subject is predictable. 
Furthermore, in this study the emotional state is represented by two values, an arousal and a 
pleasure value. To measure emotions the self-assessment manikin score (SAM) are used 
\cite{bradley1994measuring}. SAMs are an easy and reliable method for measuring the emotional 
experience of humans. One-way analysis for variance (ANOVA) was performed to find out if there 
is a statistically significant difference.

\section{Model of Emotion}

\begin{figure} 
  \centering
     \includegraphics[width=0.35\textwidth]{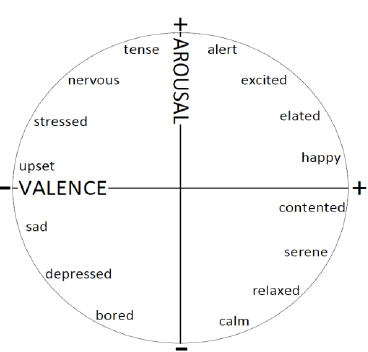}
  \caption{The circumplex model of emotion developed by James Russell \cite{russell1977evidence}}
  \label{fig:russell}
\end{figure}

Russels circumflex model of affect defines a 2D plane on which certain emotions like happiness, 
fear, sadness and anger can be placed \cite{russell1991culture, russell1977evidence}. In this model, 
arousal and pleasure are used as axis. In some cases, emotions are represented in a three-dimensional 
space, where the third dimension describes the dominance of the situation. This third dimension 
will not be considered in this work. As shown in Figure \ref{fig:russell}, the emotions happiness 
and anger have a high arousal value. However they differ in the pleasure value. Obviously happiness 
has a high pleasure value, whereas anger has a low pleasure value. The emotions of sadness and fear 
are both characterized by a low pleasure value, but differ in the arousal dimension. 
These differences in the arousal and pleasure dimension represent our hypotheses for the 
development of our questionnaire. The user study was conducted in two steps: In the first part, 
the calibration phase, the participants should assess three different situations: seeing an old 
friend after a long time, putting an animal down, (the full descriptions and questions can be 
found in the supplementary material), and the feeling he/she experiences before an important 
exam. Each situation had to be assessed by a textual description of the feelings he/she imagine to 
have in the particular situation.  The questions aim to gain an impression of how the participant 
reacts to different situations and additionally serves as a preparation phase for the second part 
of the study. In the second part, the questionnaire for each scenario consists of three elements: 
A text box and two self assessment manikin (SAM) scales, one for evaluating the arousal value and 
one for the pleasure value. Since we are focusing on modeling the emotions by the arousal and 
pleasure value, only these two values are used for the evaluation of our analysis. 
As the project is a pre-study and is supposed to find an answer to the basic question if it is 
possible to express emotions just by using special light patterns and movements of the robot itself, 
the results and the design of the study itself are not complete. This means that we have still to 
figure out how to perform the mapping of an emotional state into a color and how the transition 
between two patterns can be designed. But in order to make fast progress and to get a first impression 
of the results, we have focused on four emotions and designed the light patterns according to 
the various psychosocial emotion theories. The resulting patterns were simplified in order to 
match the low resolution of the used display. The difficulty was to figure out, how the relationship 
between the emotions and colors are, because colors can have positive but also negative effects and 
they can feel warm or cold. In another study, researchers have reported positive reactions of humans 
to bright colors and the negative reactions to dark colors \cite{kuthe2013marketing}. Similar effects 
were also significant in other studies conducted in the context architecture and marketing. 

\begin{figure} 
  \centering
     \includegraphics[width=0.25\textwidth]{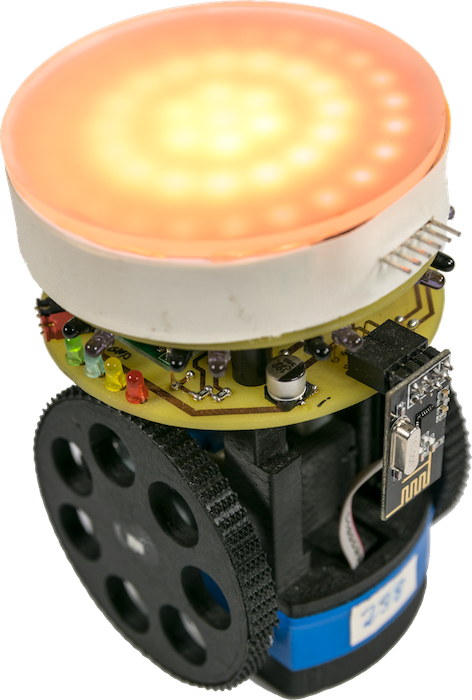}
  \caption{The two-wheeled robot with the attached low-resolution RGB-LED display.}
  \label{fig:robot}
\end{figure}

There are several color-emotion theories and models, like the "Color Theory" from Goethe and 
additionally a global overview from Claudia Cortes and Naz Kaya \cite{kaya2004relationship}. 
These models are discussed and compared in \cite{nijdam2009mapping}. Recent studies additionally 
suggest that color LEDs on a robot's face and a suitable control of them can lead to a good recognition 
rates of the displayed emotions. However, our goal is to express emotions without using an anthropomorphic 
face. Therefore, the mapping of an emotion to a suitable representation base on light patterns and 
appropriate movement patterns of a small two wheeled robot with a low-resolution LED display on top 
(see Figure \ref{fig:robot}). 
The main problem is that the interpretation of colors is subjective. The color red, for example 
is associated with both anger and joy. Also an emotion is not always associated with a specific 
color. There are cultural and personal differences, which are not subject of this work. These 
conflicts can be avoided, by using symbols which have the same meaning cross-culturally. For example, 
a rainbow is mostly a symbol for happiness, while rain may cause a depressive mood. These and 
similar facts are used together with the above mentioned color theories to create the following 
light patterns. Furthermore, besides the static color also the temporal change of the color and 
the robot movement was designed to support the expression of a particular emotion. The advantage of 
this combination is that the robot movement further improves the emotion recognition. The design of 
suitable robot movements is a separate question and is not part of this work. The combination of 
these three factors are supposed to express the emotional state of the robot, but the influence 
of each individual factor in detail is not further investigated.

\section{Emotion Display and Patterns}

In this section, the particular patterns corresponding to one of the four investigated emotions 
are described in detail. For each pattern the psychological influence of colors and their dynamical 
change as well as the blinking rhythm is described. The robot with the attached display presents each 
light pattern while driving in a gray box, separated into two rooms. The robot has a LED display 
consisting of 47 RGB-LEDs. Each LED can be individually addressed by a micro-controller adjusting 
the color, the brightness, and blink frequency. The robot starts in the smaller room of the two 
rooms, which measures 20 cm times 40 cm and subsequently moves to the bigger room which is 40 cm 
times 40 cm.

%\cite{fong2003survey}

\subsection{Fire pattern:}

This light pattern uses the effect of the color combination yellow, orange and red. These 
colors are interpreted as active, energetic and powerful. The goal is to symbolize an emotional 
state of arousal. This state can be positive or negative including fear, anger, curiosity and 
love. Since arousal is a result of stimulation, the stimulation has to be done appropriately. 
In the state of arousal the activity in mind and body is increased. An arousing emotion like 
anger is often described by the metaphor of heat, which reflects the energy produced during the 
process. The continuing impact of red can raise blood pressure and the heart rate. Keeping this 
in mind, we can try to trigger an arousal state by light. The purpose of this light pattern is 
to represent a fire. Therefore, the LEDs are switched on one after another, beginning with the 
inner circle glow yellow. In a second step, the LEDs in the middle circle glow orange following by 
the red lighting outer circle. After a short break of 3 seconds the LEDs were turned off and the 
process is repeated form the start. The break between each step is very short, so the light 
pattern appears dynamic and fast. This symbolizes the high emotional state of arousal. 
Figure \ref{fig:fire} below shows the light pattern used for expressing and triggering arousal. 
\begin{figure*}[htpb] 
  \centering
     \includegraphics[width=\textwidth]{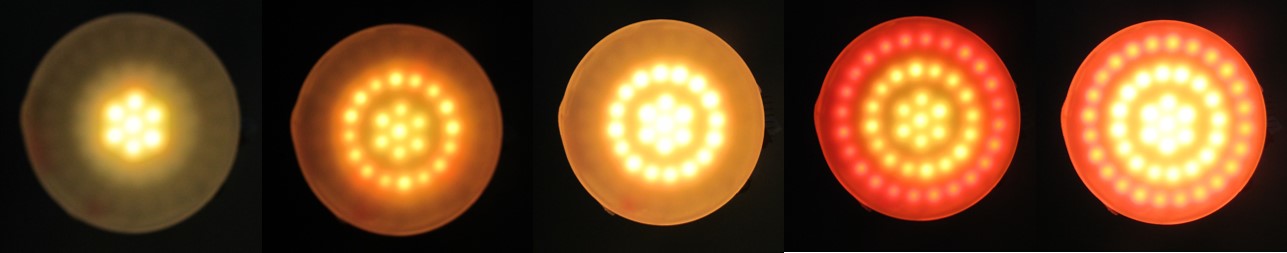}
  \caption{Fire pattern}
  \label{fig:fire}
\end{figure*}
\begin{figure*}[htpb] 
  \centering
     \includegraphics[width=\textwidth]{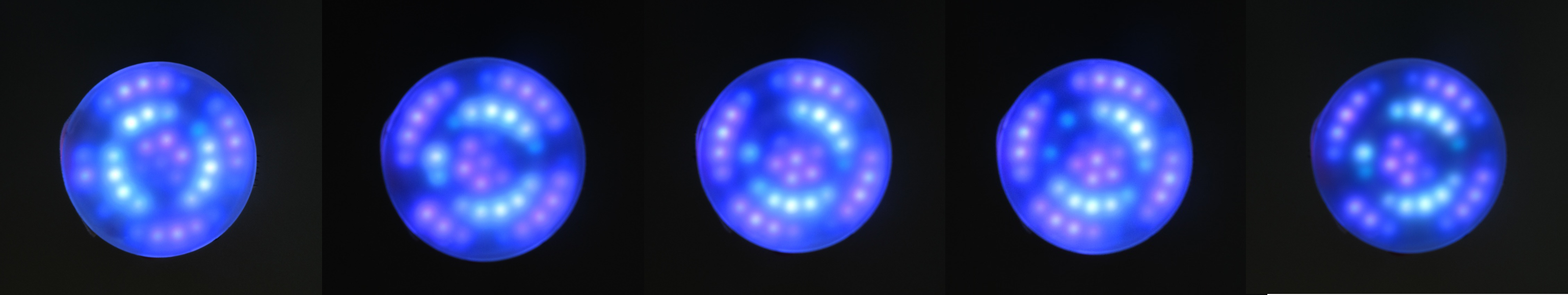}
  \caption{Rain pattern}
  \label{fig:sadness}
\end{figure*}
\begin{figure*}[htpb] 
  \centering
     \includegraphics[width=\textwidth]{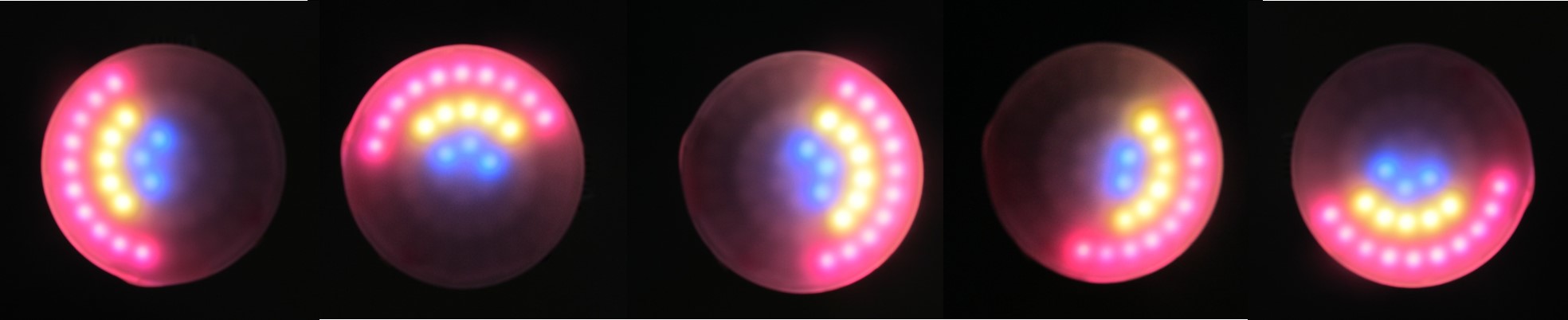}
  \caption{Rainbow pattern}
  \label{fig:rainbow}
\end{figure*}
\begin{figure*}[htpb] 
  \centering
     \includegraphics[width=\textwidth]{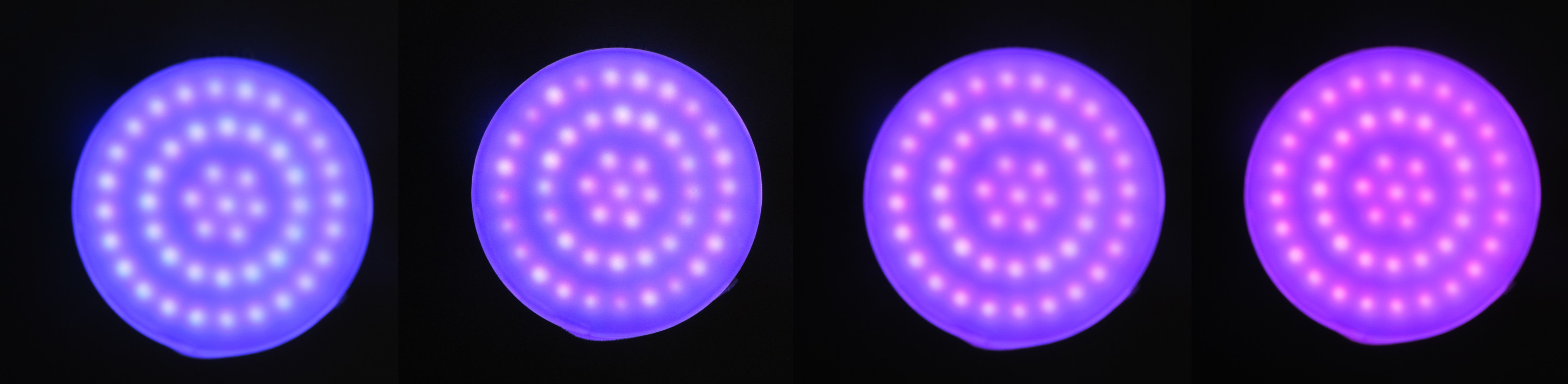}
  \caption{Purple pattern}
  \label{fig:fear}
\end{figure*}
\subsection{Rain Pattern:}
To arouse a feeling of sadness, depression and motionlessness the color combination blue-green is suitable.  
As already mentioned, there are differences in how people associate emotions with colors. Sadness is 
associated with blue in the United States and in most western countries \cite{barchard2014sadness}. In India, however it is associated with happiness. As there is no common color, we decided to model a specific blue 
pattern suitable for most western countries.

As the study of Choi et al. \cite{choi2007study} has reported, low brightness contrasts and low 
blinking rates are associated with sad mood states. Additionally, based on the fact that rain clouds 
are used in photography and animated films to introduce a depressing atmosphere, a light pattern which 
is supposed to represent a rain cloud circulating on the display was implemented. A fade in and fade 
out method was applied to strengthen this impression. Figure \ref{fig:sadness} illustrates the 
illuminated light pattern.

\subsection{Rainbow Pattern:}

Color is an integral element in the architectural environment. The goals of color design involves a 
psychological influence. The impression of color has a great importance in creating a psychological 
mood. Studies showed, that people react more positively to bright and chromatic colors (pink, red, 
yellow and blue) \cite{nijdam2009mapping} than to their complementary colors. Following this premise, 
high brightness values are associated with happiness. To increase this effect three rainbow colors, 
red, blue and yellow, which give most people a sense of happiness, are used. The brightness value 
of this light pattern is chosen relatively high, in order to suggest a more satisfying situation 
than the purple and rain pattern. Furthermore it has been proven that round shapes stand for 
peace and harmony \cite{ulshofer2010literatur}. As basic color, yellow was selected as yellow 
is correlated to pleasure and optimism in psychological studies. The color is also associated with 
the sun and summer time. The rainbow pattern is drawn step by step on the display. Every few seconds 
a LED is switched on, until all LEDs which are supposed to represent the rainbow are turned on. After 
the rainbow is completely displayed, it starts to rotate until the supporting robot movement stops. 
The rotation speed of the rainbow is relatively low, so the light pattern appears peaceful compared 
to the fire pattern. Figure \ref{fig:rainbow} illustrates the light pattern.

\subsection{Purple pattern:}

The color purple is a combination of blue and red. The color itself is neither cold 
nor warm. It communicates something mystical and depressing. Depending on the percentage of 
this two colors, the effect of the mixture is different. Lighter versions of purple, such as lavender, 
appear calm and passive. While increasing the saturation of blue, the color seems more calm and does 
not transmit an anxious state. By gradually raising the red saturation value the color is considered 
as more restless. Exploiting this fact, a light pattern was designed to express a fearful and 
frightening state. The light pattern starts with a purple color, which has a high blue value. 
After a short break, the color values change and the saturation of the blue color gets lower, 
while the saturation of red increases. These steps are repeated until the blue saturation value is 
at a minimum and the red saturation value has reached its maximal value. The saturation change is 
performed slowly, so the light pattern appears like a slow process of getting into a frighting 
situation. The purple light pattern is depicted in Figure \ref{fig:fear}.

\section{Experiment}

For evaluating the implemented patterns of emotion expression, we simulated four different scenarios. 
In each scenario, the small robot is moving in the two-room environment. The robot uses a specific 
movement pattern and light pattern, to express an artificial emotion. Based on \cite{camurri2003recognizing} 
the movement pattern in an specific emotional state can be described as follows:

\begin{itemize}
\item \textbf{fear}: frequent tempo changes with long pauses between the changes
\item \textbf{sadness}: few tempo changes, long duration and smooth transitions
\item \textbf{joy}: frequent tempo changes with longer stops between each change
\item \textbf{anger}: short duration, frequent changes of tempo with short pauses between each change
\end{itemize}

For the scenarios, which are characterized by a corresponding light and movement pattern, 
we propose the following hypotheses:
\begin{itemize}
\item \textbf{H1}: The arousal value in the situation where the sadness or fear pattern is shown, 
is lower than in situations corresponding to the feeling of happiness and anger. 
\item \textbf{H2}: There should be a significant difference between the pleasure value of the happiness 
or anger pattern, and the pleasure value of the fear or sadness pattern. 
\end{itemize}

At the beginning of each trial the robot receives a signal, which indicates the scenario. Afterwards 
it moves to the middle of the bigger room and waits for three seconds, then the corresponding 
light pattern is displayed and the robots starts to perform the corresponding movement pattern. The 
light pattern is repeatedly displayed until the robot reaches its starting point again. 

For the emotional expression of fear, the robot first rotates 90 degrees to the left, then 180 degrees 
to the right and again to the left. Between each rotation there is a pause of one second. Afterwards the 
robot drives with the highest possible speed back to his starting point in the small room, while the fear 
light pattern is shown on the LED display. 

As mentioned before, sadness is characterized by a slow tempo. Therefore, the robot drives with 25 
percentage of its maximum speed. Similarly, to the movement which is shown in the fear case, it drives 
back to the small room but in this case with a respectively lower speed. Every five seconds, the robot 
makes a pause of half a second. Furthermore, the sadness light pattern is displayed to underline the 
effect. 
 
In the joy scenario, the robot rotates as it reaches the big room. This rotation is based on a 
pirouette in human dancing.  

For the scenario, which is supposed to demonstrate an angry emotional state, the robot drives fast 
and changes its direction quite often. The corresponding light pattern is shown in parallel. \\

\noindent In the study, we had 23 participants, who range in age from 19 to 29 years. Most of them 
had a technical background. In order to avoid biases, the scenarios were shown in a randomized order 
to each participant. The textual description was served as a general verification of comprehensibility 
of the scenarios. 

\section{Results}
In order to prove the hypothesis, first the average pleasure and arousal value of each emotional 
state are compared. The scale range from 1 to 5. The average values can be found in table below. 

\begin{center}
 \begin{tabular}{lcr}
  Emotion & excitement (mean) & pleasure (mean)\\
  \hline
  anger & 4.23 & 2.36 \\
  fear  & 2.45 & 2.86 \\
  happiness & 3.64 & 3.63 \\
  sadness & 1.5 & 2.86 \\
 \end{tabular}
\end{center}

These values show that a difference between the scenarios can be observed. To be more accurate 
and prove a significant difference, an analysis of variance (ANOVA) was performed and pairwise tests 
between the different scenarios were executed. For this purpose, the scenarios with low arousal 
(fear and sadness) were grouped and compared with high excitement scenarios like anger and joy. The 
comparison reveals, that there is a significant effect (F[1,22] = 36.11, p \textless 0.001). This 
supports our first hypothesis. Unfortunately, the difference between the pleasure value in the 
situation which aims to express happiness, compared to the other scenarios was not significant. 
Thus, the second hypothesis is not fully supported by our survey. One possible reason could be that 
a high activity is easy to demonstrate by an appropriate high blink frequency, but a commonly 
accepted color interpretation is hard to observe.

\section{Conclusion}

The main gaol of this study was to evaluate, if emotion expression can be done by only using 
colors, dynamic light and motion patterns. The four developed patterns are based on psychological 
insights, like the impact of colors and rhythm. Our hypothesis was, that these emotions should be 
recognized by a human. For this purpose, different movement patterns within a small test setup were 
created and supported by appropriate light patterns, so that a particular emotional state should be expressed 
by the robot. A user study was conducted to determine if the shown scenarios can be interpreted 
correctly. On the data, an ANOVA was performed for the arousal and pleasure value of the appraisals 
of individual patterns. A statistically significant difference was observed among the group of patterns 
with high arousal value and those with a low value. This result indicates that our patterns are able to 
represent one dimension of the considered emotions and that they can be recognized appropriately in the 
case of emotions with different arousal values. However, to express emotions of pleasantness and 
unpleasantness solely on the basis of the used LED display and the corresponding movement pattern 
was difficult to distinguish by the participants. One possible reason is that the frequencies of 
the illuminated patterns are too similar. 

Further research is required to achieve clear emotion expression. Due to hardware restrictions of 
the robot drives, the driving speed was hardly varying which limited the dynamic of the movements. 
Increasing the smoothness and the maximal driving speed of the robot would lead to a more distinct 
emotion expressions.

% trigger a \newpage just before the given reference
% number - used to balance the columns on the last page
% adjust value as needed - may need to be readjusted if
% the document is modified later
%\IEEEtriggeratref{8}
% The "triggered" command can be changed if desired:
%\IEEEtriggercmd{\enlargethispage{-5in}}

% references section

% can use a bibliography generated by BibTeX as a .bbl file
% BibTeX documentation can be easily obtained at:
% http://www.ctan.org/tex-archive/biblio/bibtex/contrib/doc/
% The IEEEtran BibTeX style support page is at:
% http://www.michaelshell.org/tex/ieeetran/bibtex/
\bibliographystyle{IEEEtran}
% argument is your BibTeX string definitions and bibliography database(s)
\bibliography{emotion_display}

\end{document}